\documentclass[preprint,pre,amsmath,amssymb, showpacs]{revtex4}
\usepackage{graphicx}
\usepackage{dcolumn}
\usepackage{bm}
\usepackage{epstopdf}
\usepackage{float}
\begin{document}

\preprint{}

\title{Extensional Instability in Electro-Osmotic Microflows of Polymer Solutions}
\author{R.M. Bryce}
\email{rbryce@phys.ualberta.ca}
\author{M.R. Freeman}
 \email{mark.freeman@ualberta.ca}
\affiliation{
Department of Physics, University of Alberta, Edmonton T6G 2G7, Canada.\\
National Institute for Nanotechnology, 11421 Saskatchewan Drive, Edmonton T6G 2M9, Canada.
}

\date{\today}

\begin{abstract}
Fluid transport in microfluidic systems typically is laminar due to the low Reynolds number characteristic of the flow. The inclusion of suspended polymers imparts elasticity to fluids, allowing instabilities to be excited when substantial polymer stretching occurs. For high molecular weight polymer chains we find that flow velocities achievable by standard electro-osmotic pumping are sufficient to excite extensional instabilities in dilute polymer solutions. We observe a dependence in measured fluctuations on polymer concentration which plateaus at a threshold corresponding to the onset of significant molecular crowding in macromolecular solutions; plateauing occurs well below the overlap concentration. Our results show that electro-osmotic flows of complex fluids are disturbed from the steady regime, suggesting potential for enhanced mixing and requiring care in modeling the flow of complex liquids such as biopolymer suspensions.
\end{abstract}


\maketitle

\section{Introduction}

It is known that the inclusion of polymers into fluids allows flow fields and instabilities associated with elastic effects to be excited in pressure or other mechanically-driven liquids \cite{Bird}. Here we show that by adding dilute amounts of inert polymer to liquids extensional instabilities can be excited in microchannels under steady  electro-osmotic flow (EOF) pumping. 

For mechanically-driven liquids there is always a significant shear component to the fluid flow, due to no-slip boundary conditions; the effective slip boundary conditions of EOF suppresses the shear allowing elongational effects to be emphasized and studied. Considerable efforts have considered ``infinitely dilute" single-chain behavior of DNA in elongation flow (see, for example, Refs. \cite{Juang2004, Randall2004}), making use of electro-phoresis which places a body force on the polymer coil dragging it through a stagnant liquid. We consider here finitely dilute solutions driven at the boundary and undergoing elongtional stress in order to investigate the possibility of instabilities in such flows, which we determine can be easily excited for low viscosity solvents.

Placing polar liquids in contact with ionizable surfaces produces an electric double layer (EDL) where ions of opposite charge are attracted to the wall \cite{Ghosal}. This electrically non-neutral EDL sheaths surfaces, allowing electric fields to produce flows by driving the sheath, which in turn viscously couples to the bulk fluid away from the wall \cite{Ghosal}. Such electro-osmotic flow is desirable for use in micro- and nano-fluidics, as flow with flat velocity fronts is observed away from the EDL in uniform channels of arbitrary cross section, with shear confined to the EDL regions. As a result scalability is superior to pressure-driven flows which become difficult to drive through very narrow channels. 

Microfluidic devices usually display creeping laminar flows \cite{umixreview}, for both pressure and electro-osmotic pumping, as the achievable Reynolds numbers are small when working with small volumes in microscaled (O(100) $\mu$m typical) channels. The low Reynolds regime prevents instabilities and turbulence which enhance mixing of chemical reactants allowing timely and complete reactions. As counter examples to typical ``smooth" low Reynolds number flows see, for example, Ref.~\cite{exp_bend} where high velocities result in intermediate Reynolds numbers (O(1-100)) allowing inertial effects to induce Lagrangian chaos and efficient mixing in planar 2D microchannels and Ref.~\cite{herringbone_mixer} where bas-relief (3D microchannels, requiring more demanding microfabrication) of a staggered herringbone shape allows Lagrangian chaos and effective mixing for Re $<$ 1. 

Polymer coils in solution can be stretched in flows. Thermal agitation causes a relaxation back towards the globule coil state and polymers act as entropic springs \cite{Bird}. Polymeric solutions therefore have a relaxation time, $\lambda$, and if flow times, $t_{\textrm{f}}=L_{\textrm{Char}}/v$, where $L_{\textrm{Char}}$ is a characteristic channel dimension and $v$ is the flow velocity, are driven to be shorter than the relaxation time $\lambda$ significant elastic stress can develop and lead to secondary flows and instabilities (see Chapter 2 of Ref. \cite{Bird} for a general overview). The Deborah number, $De=\lambda /t_{\textrm{f}}$, is a dimensionless parameter describing viscoelastic solutions; when this number reaches a geometry-dependent threshold, roughly of order unity, instabilities often occur \cite{Bird}. For example, it has been shown that chaotic instabilities can arise in Dean flow of polymer solutions through curvilinear serpentine microchannels \cite{microflow}, where earlier characterization of macroscopic-equivalent flows \cite{nature} demonstrated the presence of elastic turbulence \cite{seminal}. In addition to these flows were elastic turbulence and chaotic flow are established it is known that rapid constriction/expansions can also lead to instabilities \cite{Bird}, for example modification of serpentine microchannels \cite{microflow} to add asymmetric corrugation (triangular constrictions followed by sudden expansions along the channel) demonstrates the formation of unstable vortices \cite{polymer_telsa_device} when polymeric fluid is pumped through the channel at sufficiently high rates. These unstable vortices are excited in linear channels as well and display apparent random growth and decay dynamics \cite{polymer_diode} which, due to the corrugation asymmetry, display differing dynamics depending on flow direction.  The unstable vortices arise at Re $<$ 1 for highly elastic polymeric liquids; in the same device Newtonian flow (water) also displays vortices at intermediate Reynolds numbers (Re $\approx$ 5) due to inertial effects.

\section{Experimental Setup}

\begin{figure}[!h]
\includegraphics[scale=0.3]{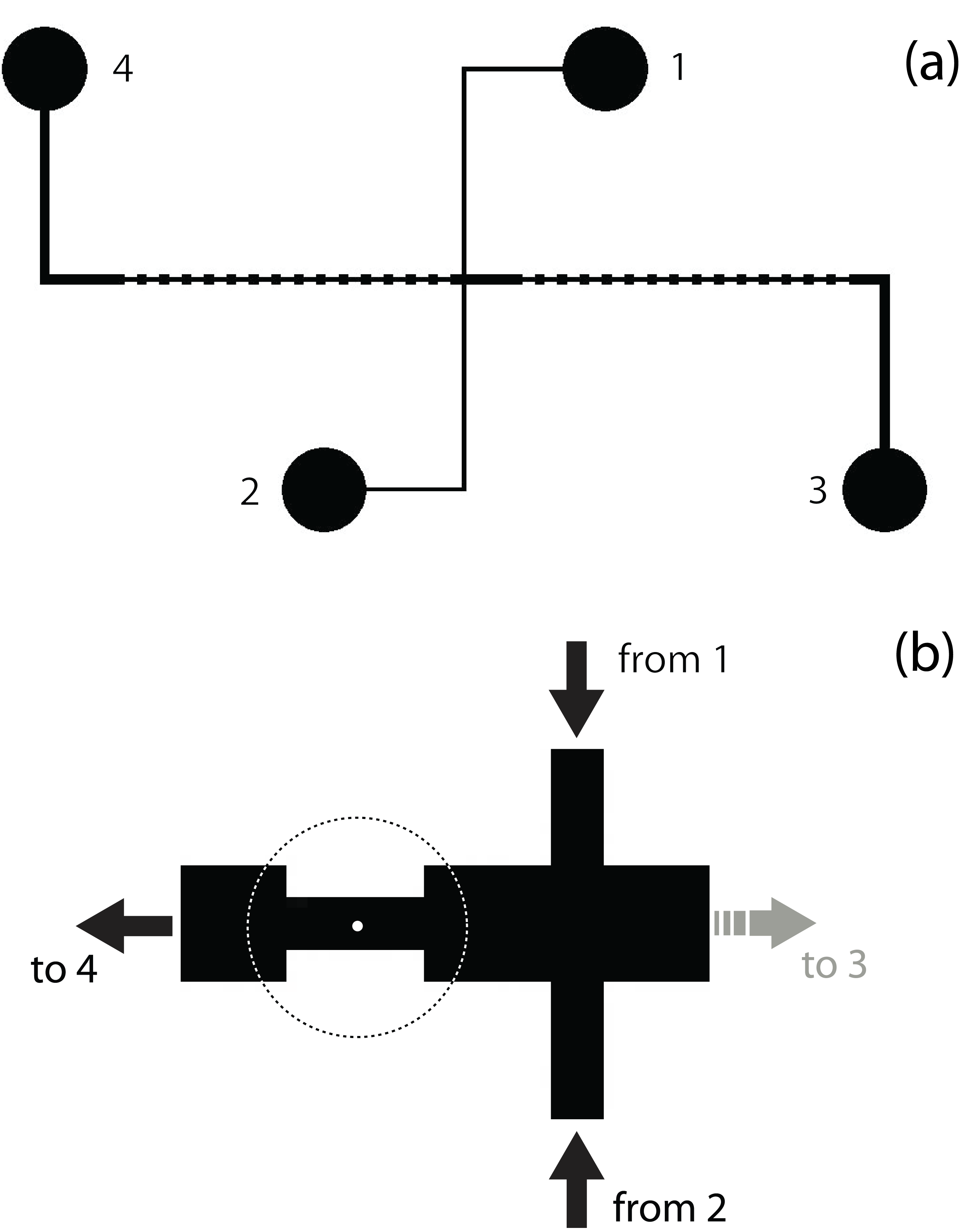}
\caption{\label{fig:Fig_schematic} Device schematic. The microfluidic device (a) consists of four-reservoirs (two input reservoirs at positive voltage that flow into the grounded fourth reservoir, the third reservoir floats during experiments described here) connected via microchannels. A dyed (reservoir 1) stream merges with an undyed (reservoir 2) stream and flows towards the ground (reservoir 4) through microchannels defined by a 20 $\mu$m deep chemical etch which results in the distinctive D-scooped profile associated with wet etching. Past the intersection that merges the two input streams 200:100 $\mu$m wide channel constrictions occur creating extensional flow fields. The small (not to scale) circle in (b) indicates the photomultiplier measurement spot ($\approx$ 10 $\mu$m focused waist of a 532 nm diode laser) used to measure fluorescence emitted by excited dye, while the large dashed circle outlines the field of view for microphotographs.}
\end{figure}

Microchannels were formed in glass with 100 $\mu$m and 200 $\mu$m width and 20 $\mu$m depth to the bottom of the ``D''-shaped chemically etched cross sections. Forming 2:1 constrictions (see Fig.~\ref{fig:Fig_schematic}.a and b) causes flow acceleration and deceleration as liquid transverses through the constricted region, inducing an extensional (stretching) flow field. Polymer solutions were made by adding $18\times10^6$ Da high molecular weight (HMW) or $5\times10^6$  Da low molecular weight (LMW) polyacrylamide (PAAm) from Polysciences Inc. to a 20:80 vol.\% methanol:water mixture, where methanol is used to prevent aging (chain scission) during storage \cite{PEOdecay}. Neighboring polymer coils impinge on each other at the overlap concentration $c_{*}$ of $\approx$ 300 ppm and $\approx$ 850 ppm for the HMW and LMW samples, respectively. We find concentrations both below and mildly above the overlap concentration of the HMW sample display similar behavior (as seen in Fig.~\ref{fig:flux_growth}.b; see also Fig.~\ref{fig:PAA_c_sweep}).

The glass microchips are loaded into a Microfluidic Toolkit \footnote{Supplied by Micralyne, Inc. For a description of the device see: Ref. \onlinecite{uTKManual}, also Ref. \onlinecite{uTKCrabtree} } and flow is electro-osmotically driven with voltages of up to 0.4 kV (electric fields as high as $\approx$ 900 V/cm). The interface between the two input streams is observed by fluorescently tagging one input solution with a small amount ($\approx 21 ~\mu$M) of tetramethylrhodamine (TAMRA) dye and exciting with a 532 nm (green) diode laser. Laser-induced fluorescence profiles are collected for various flow rates by focusing the laser to 1/10th of the channel width (10 $\mu$m beam waist) close to the dyed-undyed fluid interface centered in an extensional region (focal spot indicated with a small circle in Fig.~\ref{fig:Fig_schematic}.b), and collecting light emitted from the dye at 570 nm (yellow) with a photomultiplier tube (PMT) after filtering with a 568 nm bandpass filter (10 nm FWHM).  

Data runs were 100 s in duration, with the first 20 s removed prior to analysis to allow startup and transient behavior to attenuate. Care was taken to thoroughly clean and rinse microchannels before use to ensure clean surfaces to prevent pH shifts or other effects from modifying flows. The cleaning protocol was to flush the channels with: water ($\times2$), $\approx$ 1 M nitric acid, $\approx$ 1 M NaOH, water ($\times2$); first by filling all the reservoirs and then pulling solutions through the channels by applying vacuum to the output reservoir (reservoir 4 in Fig.~\ref{fig:Fig_schematic}.a).

Flow velocities in EOF are given by $v=\mu_{\textrm{EOF}}E$, where $\mu_{\textrm{EOF}}$ is the electro-osmotic mobility and $E$ the applied electric field. By timing the movement of a marker of dye over a known distance the electro-osmotic mobility was measured to be $(5.6\pm0.5) \times10^{-4}~\textrm{cm}^2/\textrm{Vs}$ for polymer-free solutions, which agrees with the value of $6.00\times10^{-4}~\textrm{cm}^2/\textrm{Vs}$ reported in 50 $\mu$m diameter fused silica capillaries \cite{Wright1997}. The HMW solution at 120 ppm has a $\mu_{\textrm{EOF}}$ of $(5.7\pm0.7) \times10^{-4}~\textrm{cm}^2/\textrm{Vs}$, indicating that for dilute solutions of large polymers the electro-osmotic mobility is not significantly modified from the pure solvent case. For the LMW solution at 120 ppm the mobility is reduced to  $(3.1\pm0.2) \times10^{-4}~\textrm{cm}^2/\textrm{Vs}$, possibly due to polymer coils interacting with the Debye layer \cite{dragEOF} which is on the order of 100 nm thick \cite{chargeglass}. 

We found that use of a LMW sample from Polysciences Inc. which contained $\approx$ 1\% ammonium chloride stopped electro-osmotic flow, even after dialysis. As PAAm is known not to adhere to channels \cite{DynamicReview}, and $\approx$ 1\% ammonium chloride solutions in water:methanol solvent demonstrated normal flow behavior, the mechanism preventing flow in the sample which included salt is not known. One possibility is the presence of a contaminant copolymer similar to  poly-N-hydroxyethylacrylamide \cite{polyDuramide}, which adheres to side-walls and significantly reduces the electro-osmotic mobility.

\section{Instabilities in Extensional Microflows of Polymeric Solutions}

\begin{figure}[!h]
\includegraphics[scale=0.275]{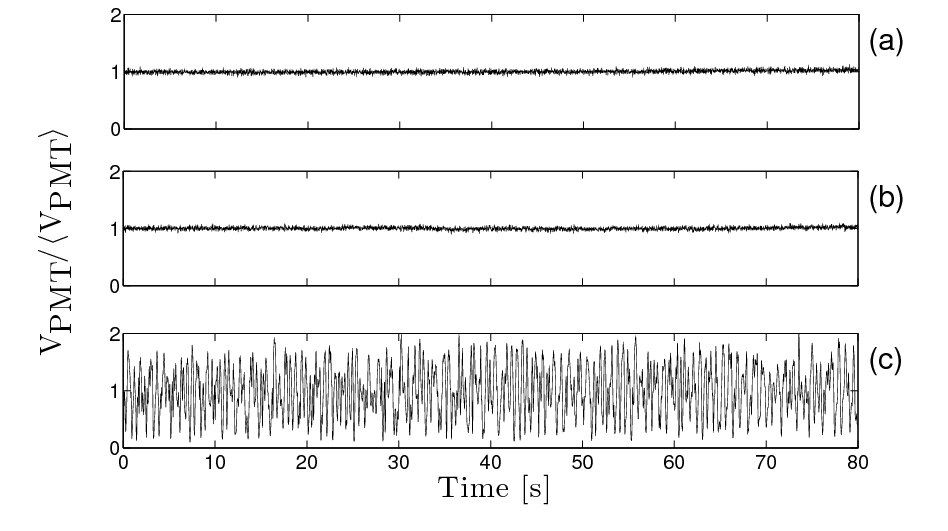}
\caption{\label{fig:Fig_PMTtimetraces} Normalized photomultiplier tube time traces. A PMT monitors the interface of the dyed/undyed streams of methanol-water solutions with (a) no driving voltage (no flow), (b) driven at 0.4 kV, and (c) driven at 0.2 kV with 120 ppm of HMW polymer added. With no polymer the PMT trace of the flow is smooth (b), albeit with noise from inherent laser diode instability as also seen for no flow (a), indicative of a stable interface and the laminar flow typical in microchannels. With polymer added rapid interface motion and instabilities are observed at moderate driving voltages, as evident by the dramatic fluctuations in the PMT signal (c).}
\end{figure}

Creating a visually observable fluid interface by flowing dyed fluid from reservoir 1 and undyed fluid from reservoir 2 to reservoir 4, Fig.~\ref{fig:Fig_schematic}.a, allows us to investigate fluid flow properties. The channel geometry is designed with 2:1 constrictions in order to induce extensional flow and the fluid interface is monitored in the center of this region (at the white spot in Fig.~\ref{fig:Fig_schematic}.b). The Reynolds numbers here are $< 0.01$, below the nominal creeping flow limit of $Re < 1$. Without polymer creeping laminar flow persists to the highest applied voltages (Fig.~\ref{fig:Fig_PMTtimetraces}.b), with the PMT traces appearing similar to the no-flow case (see Fig.~\ref{fig:Fig_PMTtimetraces}.a). 

In contrast, very large fluctuations arise, at sufficiently large driving electric fields, when polymer is added to the solution (Fig.~\ref{fig:Fig_PMTtimetraces}.c). This arises from an instability of the flow causing the local dye concentration to vary under the green excitation spot. The presence of polymer results in large movement  of the dye/undyed interface relative to the illumination spot, generating order-unity fluctuations in Fig.~\ref{fig:Fig_PMTtimetraces}.c. We note that extensional flow also exists in the cross-channel region, due to acceleration, which excites instabilities in polymeric fluids at high flow rates; the Deborah number in the cross-channel region is lower than in the designed constriction and these instabilities are eliminated at low flow rates.

\begin{figure}[!h]
\includegraphics[scale=0.28]{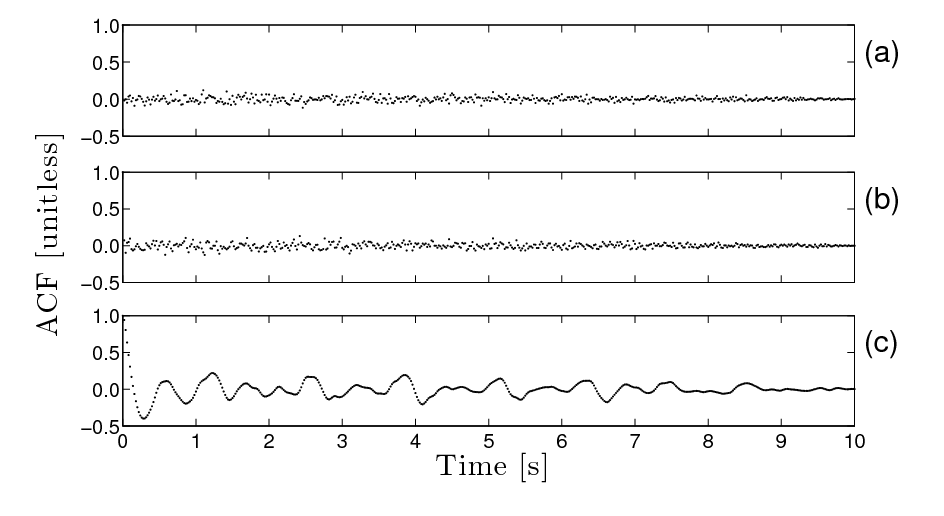}
\caption{\label{fig:ACF_PMTtraces}  Autocorrelations of photomultiplier tube signals (Fig.~\ref{fig:Fig_PMTtimetraces} data). Methanol-water solutions with (a) no driving voltage (no flow), (b) driven at 0.4 kV, and (c) driven at 0.2 kV with 120 ppm of HMW polymer added. The autocorrelations of polymer free fluid flow decorrelate within a few sampling time steps and no substructure is apparent, while unstable polymer flows show rapid drop off followed by substructure without prominent peaks, indicating the presence of a multitude of frequencies. }
\end{figure}

\begin{figure}[!h]
\includegraphics[scale=.75]{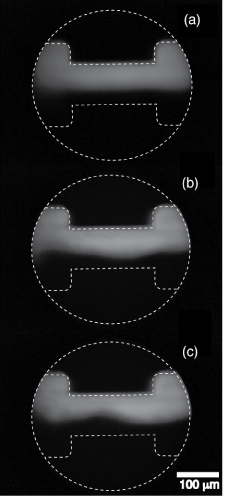}
\caption{\label{fig:photographs}  Photomicrographs of (a) no polymer and (b, c) 120 ppm HMW polymer flows at 0.2 kV.  Image (c) is taken one second after (b). Dashed lines indicate field of view and microchannel constriction outline. The interface between dyed and undyed liquid streams is (a) stable for polymer free samples and becomes (b, c) unstable with the inclusion of polymer. The scale bar is 100 $\mu$m.}
\end{figure}

Photomicrographs (Fig.~\ref{fig:photographs}) captured by wide field excitation of the dye using an external laser show the stable interface found in polymer-free flows (Fig.~\ref{fig:photographs}.a), sampling this interface with a focused laser spot and a PMT results in a smooth time trace (see Fig.~\ref{fig:Fig_PMTtimetraces}.b). Adding polymer leads to fluctuations, of which the largest scale features are apparent in Fig.~\ref{fig:photographs}.b and c (compare with Fig.~\ref{fig:Fig_PMTtimetraces}.c). The largest spatial scale of viscoelastic instabilities is determined by the system size, as for inertial instabilities \cite{Bird}. There is no clear lower bound on spatial scale, with polymer size, separation, or diffusion lengths being possible cutoff lengths \cite{gs_review2004}.

No dominant oscillations were observed in PMT signals collected. The autocorrelation of PMT traces of polymer solution flows have quick decay times ($\approx$ 0.1 s decorrelation time, taken as the e-folding time), and displayed substructure indicative of multiple frequency components while not exhibiting sharp, large, repeated ringing which denote dominant sinusoidal frequencies. Autocorrelations of PMT signals for polymer-free fluid decorrelated in $\approx$ 2 sampling time steps and showed no substructure. Fig.~\ref{fig:ACF_PMTtraces} shows the autocorrelation function for Fig.~\ref{fig:Fig_PMTtimetraces} data. Scatter and spectral plots (not shown) also did not show evidence of distinct sinusoidal frequencies \cite{ebook} in unstable polymer flow. Taken together these measures indicate that the flow is chaotic \cite{microflow}. More detailed evidence is required in order to determine if the system investigated here displays true elastic turbulence \cite{gs_review2004}, for example measurement of flow velocity fluctuations and their spectral scaling properties.

The PMT signal is dependent on dye concentration, possible thinning effects \cite{Arratia}, and flow velocities. To reduce such effects the PMT signal is normalized by the average voltage. To study the development of the instability versus driving voltage, the PMT signal fluctuations, $\tilde{ \delta}_{\textrm{PMT}} $, defined as $\textrm{var}\left (\frac{\textrm{V}}{\langle \textrm{V} \rangle} \right )$, where var(V) is the variance of the PMT voltage signal from the average  $\langle \textrm{V} \rangle$, are measured for various driving potentials. The precise location of the excitation spot from the interface will affect our measure $\tilde{ \delta}_{\textrm{PMT}}$. Our absolute positional accuracy is roughly 5 $\mu$m and, by measuring away from the interface on either side for a given experimental alignment, we estimate the introduced error to be roughly 10\% of the measured fluctuation for unstable flows at higher flow rates (0.2 kV and above driving voltages) which is also the estimated error found by comparing differing experimental runs (which requires realignment). At lower flow velocities the system noise begins to dominate, and for no-flow conditions run-to-run differences in measured fluctuations is ca. 30\%.

\begin{figure}[!h]
\includegraphics[scale=0.255]{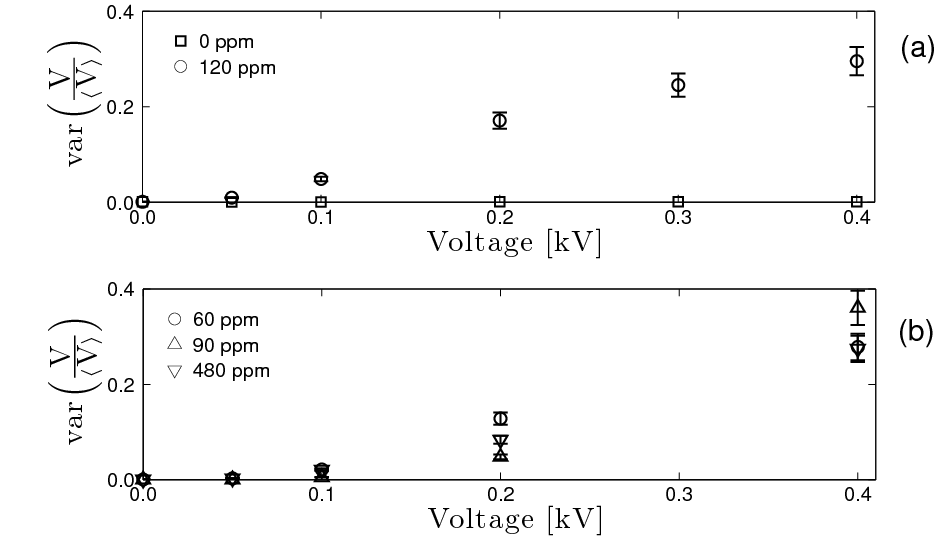}
\caption{\label{fig:flux_growth}  Fluctuation amplitude in normalized fluorescence signals. Fluctuations in the motion of the dyed/undyed fluid interface are measured with normalized PMT voltage for (a) 120 ppm of HMW polymer; the error is roughly on the order of 10\% at higher flow rates for unstable flows. Control (polymer-free) solution fluctuations are nominally zero (compare with Fig.~\ref{fig:Fig_PMTtimetraces}.b). The instability threshold is predicted to be $\approx$ 0.02 kV. Similar behavior is displayed (b) for 60, 90, and 480 ppm samples of HMW polymer.}
\end{figure}

Polymer-free solutions display laminar flow for all driving voltages (Fig.~\ref{fig:flux_growth}.a). In contrast to this laminar flow, fluctuations grow in size as the driving voltage increases for a concentration of 120 ppm of the HMW sample (Fig.~~\ref{fig:flux_growth}.a). As seen in Fig.~~\ref{fig:flux_growth}.b similar behavior occurrs at other concentrations of the HMW sample, both below (60, 90 ppm), and above (480 ppm) the overlap concentration.

\begin{figure}[!h]
\includegraphics[scale=0.28]{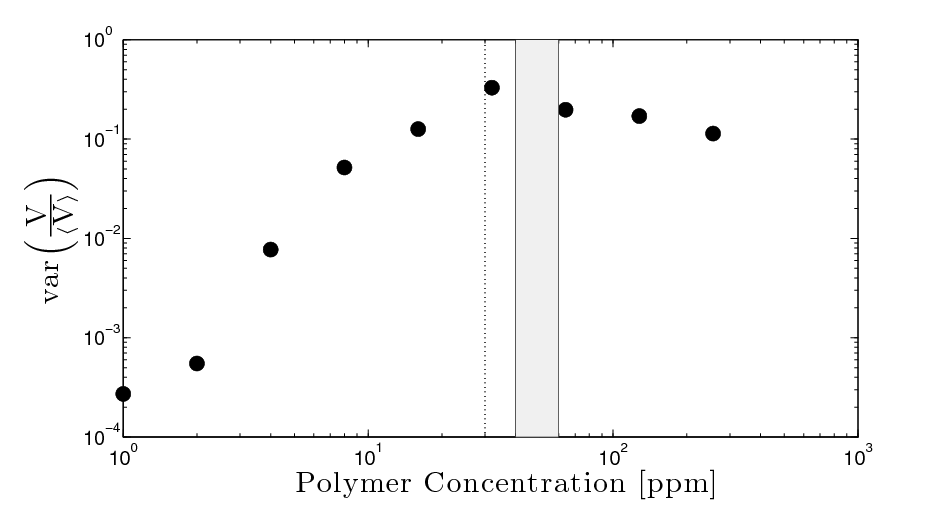}
\caption{\label{fig:PAA_c_sweep}  Fluctuations at various concentrations. The fluctuations in normalized PMT voltage for HMW polymer at different concentrations and 0.2 kV. It can be seen that fluctuations increase with concentration but rapidly peak and plateau well below the overlap concentration ($\approx$ 0.1$c_{*}$; $c_{*} \approx 300$ here). Modeling the polymeric solution as a hard colloidal suspension predicts a transition in behavior at c $\approx$ 40-60 ppm (grey band) due to molecular crowding. Simulations of a similar polymer \cite{dilutenotdilute} find a threshold of 0.1$c_{*}$ (dotted line) above which hydrodynamic interactions between polymer coils significantly affects properties. The good agreement of both the colloidal model and the simulations with the observed peaking of fluctuation growth suggests hydrodynamic interactions and molecular crowding is the mechanism inhibiting fluctuations.}
\end{figure}

Scaling properties of viscoelastic instabilities are not well understood, but the basic elements of curvature and elastic normal stress (stretching) along streamlines are well established as underlying instability \cite{ei_scale} as these mechanisms lead to ``hoop-stress" \cite{Bird} allowing stretched polymers to cross streamlines, disturbing laminar flow. Fluctuations should grow with concentration (as the movement of individual polymer coils underlies the fluctuations). Changing the concentration and investigating fluctuations at a constant (0.2 kV) voltage indicates that fluctuations rapidly grow from the noise floor and then peak and plateau with a gentle decrease as we increase concentration, with fairly uniform measured fluctuations from $\approx \frac{1}{10}$X to $\approx$ 1X the overlap concentration, Fig.~\ref{fig:PAA_c_sweep}.

Our fluctuation measure $\textrm{var}\left (\frac{\textrm{V}}{\langle \textrm{V} \rangle} \right )$ takes its maximum possible value of 1 for square-wave signals; the precise maximal value for a given signal is $\leq 1$ and depends on the nature of the waveform. Smooth growth in our measure occurs until the maximum, for a given waveform with linear amplitude growth. When the maximal value is reached a sudden apparent threshold will occur and our measure will saturate. The maximal threshold observed here is followed by a mild decrease rather than a strict capping and saturation; this is suggestive of a qualitative change in behavior rather than an artifact of our measure.

Modeling the polymer solution as a colloidal solution \cite{p_colloid} one expects hydrodynamic interaction mediated molecular crowding effects above a volume fraction $\phi$ of approximately 0.1 \cite{colloid_text} where such effects are empirically found \footnote{The presence of molecular crowding in colloids was first noted by a striking increase in the rate of viscosity growth with concentration due to hydrodynamic interactions leading to particle-particle interaction, resulting in viscosity ``hardening" \cite{colloid_text}.}. To estimate the concentration when hydrodynamic interactions between polymers play a significant role we take $\frac{c_{thresh}}{c_{*}}=\frac{\phi_{thresh}}{\phi_{*}}$, with $\phi_{thresh}$ empirically known \cite{colloid_text} to be $\approx$ 0.1 and $\phi_{*}$ corresponding to the overlap concentration which, depending on the particular packing taken near overlap, ranges from roughly 0.5-0.7; this leads to an estimate of $c_{thresh} \approx$ 40-60 ppm for our HMW polymer; this estimated threshold is close in value to the observe peak in Fig.~\ref{fig:PAA_c_sweep}. The similarity in thresholds suggests that crowding may be the mechanism inhibiting growth in measures fluctuations; the tendency for polymer coils to deform makes the hard colloid model, which assumes polymer coils act as hard spheres, only approximate. It is known that molecular crowding strongly affects and reduces molecular transport \cite{mcrowd} by hindering movement of macromoleclues; hindered movement of polymers across streamlines will limit fluctuations, as we observe here. 

Compelling evidence that hydrodynamic interactions are important also comes from brownian dynamic simulations of $\lambda$-DNA, a biopolymer similar in size and nature to our HMW sample \cite{microflow}, in shear and elongational flows \cite{dilutenotdilute} which demonstrate that for $\frac{c}{c_{*}}$ above 0.1 including hydrodynamic interactions significantly affects calculated properties. The excellent agreement of the $\approx$ 0.1$c_{*}$ threshold found in the simulation of (bio)polymers where hydrodynamic interactions become important and the peaking of fluctuations here at $\approx$ 0.1$c_{*}$ supports the hypothesis that hydrodynamic interactions are limiting growth in fluctuations. 

We note that below the predicted threshold, above which hydrodynamic interactions have significant affect on bulk properties, our measure has approximate (square) power-law scaling as expected for the variance given linear growth in amplitude, which is in turn is expected if polymer coils are independent. The approximate power-scaling occurs with a concave rounding versus the strictly linear curve predicted on a log-log plot. This rounding (breaking of power-law scaling) suggests subtle polymer-polymer interactions even in quite dilute solutions; recent systematic and careful rheological studies have demonstrated \cite{diluteNOT} polymer-polymer interactions can exist for concentrations as low as 1\% of the overlap concentration in elogational flows.

The lack of true power-law scaling at low concentrations, the mild decrease in fluctuations above a threshold, and the correspondence of this observed threshold with the predicted threshold for hydrodynamic interactions in (hard) colloids and (soft) polymers indicate hydrodynamic interactions are important in nominally dilute solutions and limit fluctuation growth.

The relaxation time for our polymers is estimated with \cite{p_dilute, diluteNOT, Colby03}

\begin{equation}
\lambda_{\textrm{Zimm}} \approx \frac{\eta_{\textrm{s}} R_{\textrm{g}}^3}{k_{\textrm{B}}T}.
\label{eq:Zimm}
\end{equation}

Here $\lambda_{\textrm{Zimm}}$ is the Zimm relaxation time estimate \footnote{In the ultradilute limit various precise prefactors $<$1 are often used in the Zimm estimate, however an unity prefactor has long been used in estimates \cite{p_dilute, Colby03} and for nominally dilute solutions recent systematic rheological \cite{diluteNOT} experiments demonstrate that the $\approx$ 1 prefactor describes relaxation time in elongational flows. FENE dumbbell models, a simple model that takes finite extension into account and offers reasonable agreement with experiments, gives a similar estimate, again with a prefactor of $\approx$ 1, for example see equation (13.5-30) and discussion in Ref.~\cite{Bird}. There is growing evidence that ``dilute" solutions must be separated into dilute (below overlap in a stagnant liquid, but with possible hydrodynamic interactions, and even steric interactions due to coil stretching, in flowing fluids) and ultradilute \cite{ultradilute} (displaying true molecular independence, even in flows) regimes.} and $\eta_{\textrm{s}}$ the solvent viscosity that immerses the individual polymer coils of radius of gyration $R_{\textrm{g}}$. The solvent frictionally slows down relaxation which is itself driven at a given thermal energy $k_{\textrm{B}}T$ which sets the level of agitation causing polymers to collapse into a random globular coil.

It should be noted that in studies of viscoelastic fluids and their instabilities high viscosity solvents are most often used \cite{Bird}, as this both increases the relaxation time due to the scaling with $\eta_{\textrm{s}}$ and reduces the effects of diffusion which will tend to reduce contrast if dye is used for visualization. For EOF driven fluids velocity will scale inversely with $\eta_{\textrm{s}}$, the effect of increasing viscosity will therefore both increase relaxation time \emph{and} reduce the achievable flow rates in such a manner that the effects will cancel leaving the De number unchanged. We initially attempted flows of high viscosity solvents ($\approx$ 65 \% sugar and equivalent glycerine), however we found it difficult to work with such flows due to long start up times and low achievable velocities making it difficult to study the flows. Joule heating effects further complicate highly viscous flows. As low viscosity solvents are most often used for microfluidic applications, De is not expected to change in a significant manner with changing viscosity for EOF flows, and experimental studies were complicated by the use of high viscosity solvents we moved to our low viscosity solvent.

Dynamic Light Scattering experiments (Brookhaven Instruments 200SM) for 120 ppm of the HMW sample gives a hydrodynamic radius of 296 nm while the LMW sample at 120 ppm has a radius of 137 nm. Using the experimental relationship \cite{Patterson} between the radius of gyration $R_{\textrm{g}}$ and the hydrodynamic radius $R_{\textrm{h}}$ for polyacrylamide ($R_{\textrm{h}}=0.68R_{\textrm{g}}$; close to the Kirkwood-Riesman prediction \cite{Kok81} of $R_{\textrm{h}}=0.665R_{\textrm{g}}$) allows us to determine the relaxation times of 0.025 and 0.0025 s for the HMW and LMW samples, respectively. Using the Mark-Houwink parameters \cite{Orwoll1999} for PAAm in pure water gives relaxation times of 0.028 and 0.0029 s. The similarity in predicted relaxation times indicates that the addition of 20\% MeOH has only a mild effect on the radius of gyration. We take the relaxation times as 0.03 and 0.003s.

The viscoelastic instability condition (see section 2.8 in \cite{Bird}, see also \cite{coil_transition_elongational}) is given by 
\begin{equation}
De=\frac{\lambda v}{L_{\textrm{Char}}} 
          \simeq  \frac{\lambda_{Zimm} v}{L_{\textrm{Char}}}
          =\frac{\lambda_{Zimm} \mu_{\textrm{EOF}} E}{L_{\textrm{Char}}} > \frac{1}{2}.
\label{eq:De}
\end{equation}
That is, electric fields give rise to extensional flows \cite{Randall2004}, with significant coil stretching, when $De>1/2$ \cite{Bird, coil_transition_elongational} for dilute solutions; note that for melts a threshold of 0.5 is also reported, which indicates that this approximate threshold value may be universal \cite{Malkin_rheology}. For the HMW sample we calculate $V_{\textrm{cr}} \approx 0.015$ kV for our system \footnote{The entrance length at a sudden contraction for creeping pressure driven flows is $\approx 0.63D$ \cite{Nguyen2002}, and has been found to be $\approx 0.57D$ for creeping electroosmotic flows \cite{Zhang2002}. Here we take $L_{Char}\approx 0.6D = 12 \mu$m.}.  Above this threshold instability is observed, however below $\approx$ 0.05 kV bias it becomes difficult to obtain data, due to diffusion rates becoming non-negligible for the low flow speeds, and the fluctuation in signal approaching the noise floor. 

\begin{figure}[!h]
\includegraphics[scale=0.265]{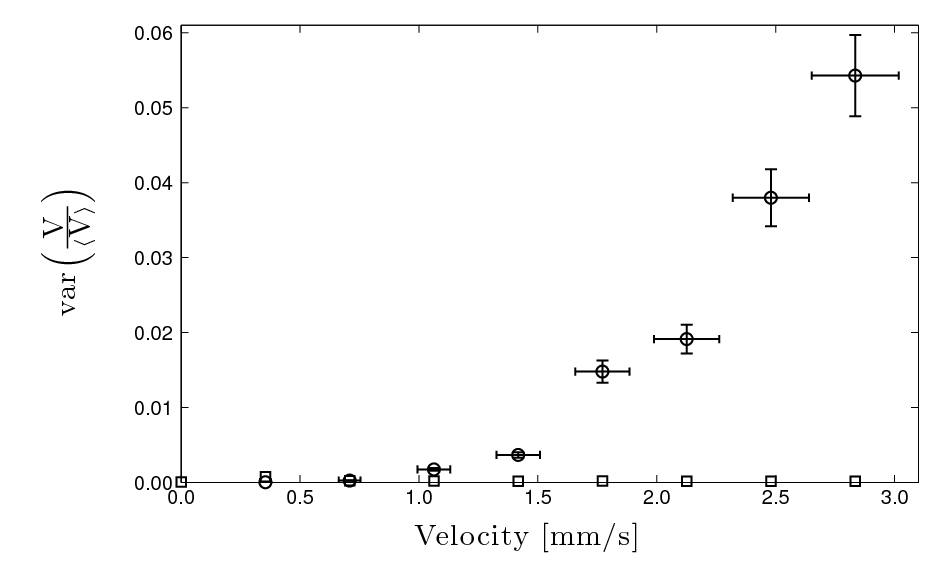}
\caption{\label{fig:flux_transition}  Fluctuation growth transition. A 120 ppm LMW sample (circles) with a predicted elastic instability at $v_{\textrm{cr}} \approx 2 ~$mm/s experimentally demonstrates an instability onset at $\approx$ 1.5 mm/s. Above $\approx$ 1.5 mm/s fluctuations grow dramatically, as compared to a gentle increase below the observed threshold; polymer free solutions (squares) display no significant fluctuations.}
\end{figure}

To better observe the transition to instability the LMW sample is used at a concentration of 120 ppm. The predicted critical driving voltage for the LMW sample is over an order of magnitude larger, $V_{\textrm{cr}} \approx 0.28$ kV (corresponding to a critical velocity of roughly 2 mm/s), than for the HMW sample. The threshold observed in Fig.~\ref{fig:flux_transition}, which occurs near the predicted value, supports the hypothesis that polymer stretching drives the instability \footnote{Comparison of Fig.~\ref{fig:flux_growth} and Fig.~\ref{fig:flux_transition} clearly reveals the fluctuations in Fig.~\ref{fig:flux_transition}  are smaller. To clarify, this is due to the De numbers being an order of magnitude smaller in the latter figure as determined by the choice in polymer molecular weights used. }. The discrepancy between the observed and predicted threshold we attribute to both the relaxation time estimate and to existence of molecular weight distribution in polymer samples. Additionally, the theoretical critical Deborah number of 0.5 is approximate, and in direct experimental observation of DNA polymers in extensional flows \cite{Chu} at a Deborah number of \footnote{In Ref. \cite{p_dilute} it is noted that liquids are made high viscosity in single DNA polymer studies (e.g. \cite{Chu}) by adding sugar, rendering the liquids highly sensitive to evaporation and temperature effects; this in turn can modify the relaxation time and possibly lead to incorrect thresholds. The 0.4 threshold observed versus the theoretical 0.5 threshold in Ref. \cite{Chu} is within the possible error introduced in this manner \cite{p_dilute}.} $\approx$ 0.4 significant stretching occurred (the latter result would reduce the estimated $V_{\textrm{cr}}$ here to $\approx$ 0.23 kV, a velocity of 1.6 mm/s). Within the limited precision of predictions our experimentally observed instability threshold agrees well with theory.

Two alternative instability mechanisms must be discussed. Conductivity gradients between co-flowing streams can give rise to electrohydrodynamic (EHD) instabilities when driven by electric fields \cite{EHD}. The possibility of EHD instability is ruled out, as the (otherwise identical) polymer free solutions do not show instability. It is also conceivable that dynamic polymer-wall adhesion \cite{DynamicReview} of polymer coils coating the walls could lead to a time varying $\mu_{\textrm{EOF}}$, driving instabilities. We reject this as underlying the instability, as PAAm does not adhere to glass \cite{DynamicReview}. This rejection is also supported by the similarity between $\mu_{\textrm{EOF}}$ measurements of polymer-free and dilute (120 ppm) HMW polymer solutions which indicates no significant modification at the wall occurs.

\section{Formation of Globules}

\begin{figure}[!h]
\includegraphics[scale=1]{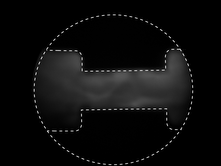}
\caption{\label{fig:globule}  Globule formation.  Globules are formed at larger applied voltages ($\geq$0.5 kV). Seen here are globules formed at 1 kV potential for a 120 ppm HMW sample. It was found that at larger applied voltages globules would form in the fluid with dimensions on the order of the channel, presumably due to cross linking of PAAm at the electrodes. While this effect may be of interest in making microgels \cite{microgel} on demand in microfluidics it is not explored further here.}
\end{figure}

We found that at large applied voltages globules would form which appeared to be gelled ÒraftsÓ of polymer (see Fig.~\ref{fig:globule}). The observed structures retained their form while the globule remained in the field of view. These globules only formed in the HMW polymer sample, at higher voltages ($\gtrsim 0.5$ kV) and $c \ge$ 60 ppm; at lower voltages, or for the LMW sample at 120 ppm, no indication of gel formation was observed. At the voltages reported in the present study ($\leq 0.4$ kV) no visual evidence of gel formation was observed in the HMW sample, and at no voltage was gellation observed for the LMW sample. 

The mechanism underlying gel formation is likely thermally activated cross-linking at the (Pt) electrode; electrogeneration of polymers at metal surfaces is an established synthesis technique which requires high current densities \cite{electropoly_PAA}. The possibility of contact glow electrolysis is discounted as current-voltage characteristics are linear both in the presence and absence of globules, which is not consistent with contact glow electrolysis where linearity is only observed pre-ignition of the plasma \cite{contact_glow_electrolysis}. Dip coating \cite{dipcoat} the electrodes and intentionally hard drying polymer films in place could either prevent flow, or resulted in globule formation at lower voltages (0.2 kV), and removal of films prevented globule formation at lower voltages. This result supports a high current density/thermal route to globule formation where pin hole or other imperfections in hard dried films would lead to a high local current density and subsequent heating effects. It is also known that polyacrylamide will spontaneously adhere to metal and form a partial film \cite{kinetics_PAA_Au}, this partial film will result in higher current densities through the non-coated Pt surface relative to an pristine surface and a mildly higher surface coverage appears to arise for higher molecular weight samples \cite{kinetics_PAA_Au}. This affinity for coating electrodes requires careful cleaning between runs \footnote{Once hard dried in place the transparent films are difficult to remove and require long hydration and careful cleaning to remove.}, while a dependence of coverage on molecular weight could account for no observed globules arising in the LMW samples for voltages observed to give rise to globules in the HMW samples. Spontaneous coating of electrodes is a rapid effect \cite{kinetics_PAA_Au}, on the order of 10's of seconds, and experimental setup times ensure that the coating process is complete before experiments commence. 

\section{Conclusion}

In summary, extensional instabilities are excited in polymer solutions electro-osmoticaly pumped through a 2:1 microchannel constriction. Polymer-free solutions display creeping laminar flow, while addition of high molecular weight polymer leads to instabilities above a critical flow rate corresponding to the viscoelastic instability condition for extensional flows. Electo-osmotic pumping is increasingly relevant as dimensions scale down and viscoelastic effects become more pronounced for small dimensions due to the inverse scaling of the Deborah number with channel size (see equation \ref{eq:De}). We observe a dependence in measured fluctuations on polymer concentration which rapidly plateaus well below the overlap concentration, occurring near the threshold where hydrodynamic interactions between polymer coils become important. At higher voltages microgels were formed at the electrodes and moved downstream. Our experimental results demonstrate instabilities can be readily excited, even in microscaled devices where laminar flow normally predominates. Electro-osmotic flow can be made unstable intentionally through the addition of small amounts of chemically inert polymer allowing the possibility for enhanced mixing. Our results also suggest caution when electro-osmically pumping biological \cite{DNA, hfilter_spit}, or otherwise viscoelastic, samples where instabilities may arise under conditions typically assumed to be laminar.

\begin{acknowledgments}
We are grateful for support from the National Institute for Nanotechnology (NINT), the Natural Science and Engineering Council of Canada, the informatics Circle of Research Excellence, and the Canada Research Chairs program. RMB is supported in part by a National Research Council Canada graduate student supplemental scholarship held at NINT. We thank CMC Microsystems for fabricating the microdevices under a CMC fabrication grant and Dr. John Crabtree of Micralyne Inc. for exemplary equipment support.
\end{acknowledgments}

\end{document}